\documentclass[11pt,twoside]{article}
\usepackage{asp2010}

\resetcounters

\begin{document}

\title{When asymmetric cosmic bubbles betray a difficult marriage: the study of binary central stars of Planetary Nebulae}
\author{Henri M.J. Boffin$^1$ \& Brent Miszalski$^2$
\affil{$^1$European Southern Observatory, Santiago, Chile}
\affil{$^2$South African Astronomical Observatory \& Southern African Large Telescope Foundation, South Africa}
}

\begin{abstract}
Planetary Nebulae represent a powerful window into the evolution of low-intermediate mass stars that have undergone extensive mass-loss. The nebula manifests itself in an extremely wide variety of shapes, but exactly how the mass lost is shaped into such a diverse range of morphologies is still highly uncertain despite over thirty years of vigorous debate. Binaries have long been thought to offer a solution to this vexing problem. Now, thanks to recent surveys and improved observing strategies, it appears clearly that a binary channel, in particular common-envelope (CE) evolution, is responsible for a large fraction of planetary nebulae. Moreover, as planetary nebulae are just ``fresh out of the oven" compared to other post-CE systems, they provide invaluable contributions to the study of common-envelope evolution and to the formation of jets in binary systems. Our studies have also started to identify strong links between binarity and morphology, including a high proportion of bipolar nebulae and rings of low ionisation filaments resembling SN 1987A. Equally important are the newly found binary CSPN with intermediate periods, which appear linked to chemically peculiar stars whose composition was modified by binary evolution. Their study may also reveal much information on mass and angular momentum transfer processes in binary stars. Here we show examples of four PNe for which we have discovered their binary nature, including the discovery of a rare case of a barium-rich cool central star. 
\end{abstract}

\section{The binary hypothesis}
Planetary nebulae (PNe) are, unlike their name suggest, the glowing gaseous envelopes ejected and illuminated by the remnants of solar-like stars on the verge of retirement as white dwarfs. They are thought to be the very brief ($\sim$\,$10^4$ years) swan song of low to intermediate mass stars leaving the asymptotic giant branch (AGB). The AGB and PN phases are the main contributors of carbon enrichment in the Universe -- the main reason that we are all ``stardust''. They recycle a significant fraction of their mass into the interstellar medium (ISM) and are thus vital probes of stellar nucleosynthesis processes and ISM enrichment. Moreover, due to their high luminosities and strong emission lines, they are important tracers of kinematics and chemical abundances in galaxies.

PNe are among the most complex, varied and fascinating of celestial phenomena. They display an astonishing variety in their shapes, ranging from spherical to highly asymmetric forms, in addition to showing complex sub-structural features of jets, filaments and rings. The spherical shape responsible for the \emph{planetary} in planetary nebula is in fact now thought to apply to only about 20\% of all planetary nebulae. A major goal over the last decades has been to understand what could be the causes of these various shapes \citep{balick_2002, demarco_2009}. The current principal contender to explain such 
 high levels of collimation and axisymmetry, which has recently been gathering support, is the binary hypothesis: that the shaping would be due to the interaction of a 
  stellar or even sub-stellar \citep{demarco_2011} companion in a binary system, either following a common-envelope (CE) phase \citep{soker_1989, sandquist_1988} or wind accretion in a wider system \citep{theuns_1996, mastrodemos}. In the first case, the common envelope will lead to a shrinking of the orbit, following spiral-in, and produce very close binaries with orbital periods of a few hours to a few days. Such systems can be detected by eclipses, ellipsoidal variations or irradiation effects -- i.e. by photometry -- or more recently by finding specific features and/or radial velocity changes in their spectra. For the second case, the systems shall be much wider and the outcome may be a symbiotic system or a barium star \citep{BoJo_1988}, with orbital periods of several hundred days. They can be detected by studying central stars apparently too cool to ionise their nebulae, when in fact the ionising white dwarf is often detectable only in the ultraviolet.
  
 Pushing even further the binary hypothesis, \citet{moe_2006} argued that most (up to 80\%) of all 1--8\,M$_\odot$ stars will \textbf{not} produce a visible PN (see also  \citeauthor{demarco_2011} 2011)!  \citet{ciardullo_2006} also had to resort to binaries to explain the existence of [O III]-bright PNe in early-type galaxies. Thus, binary interactions would not only be required to shape but also to produce PNe, as the clear, dramatic outcome of the difficult marriage between a couple of stars.
     
To confirm the binary hypothesis, it is crucial to find these binaries inside PNe and to demonstrate that they are causing the bewildering variety of PN shapes. Once a large population of binaries has been found, it will be possible to start to do comparisons with population synthesis models, and thereby constrain the numerous unknown parameters that these have introduced, in particular the efficiency of the common envelope mechanism in shrinking the orbit. 

Few binaries had been known at the centre of PNe until a few years ago, but thanks to very efficient and well-targeted surveys, we are seeing the turn of the tide. 
By cross-correlating OGLE time-resolved photometry and Bulge PNe catalogues, \citeauthor{miszalski_2009a} (2009a, 2009b) revealed a wealth of new close binary stars as CSPN and more than doubled the number of binary CSPN known! These studies provided a close binary fraction among CSPN of $17\pm5$\%, which is clearly a lower limit as the detected fraction of CSPN binaries reflects only the closest of all systems. 

\begin{figure}
	\centering
	\includegraphics[width=0.75\textwidth]{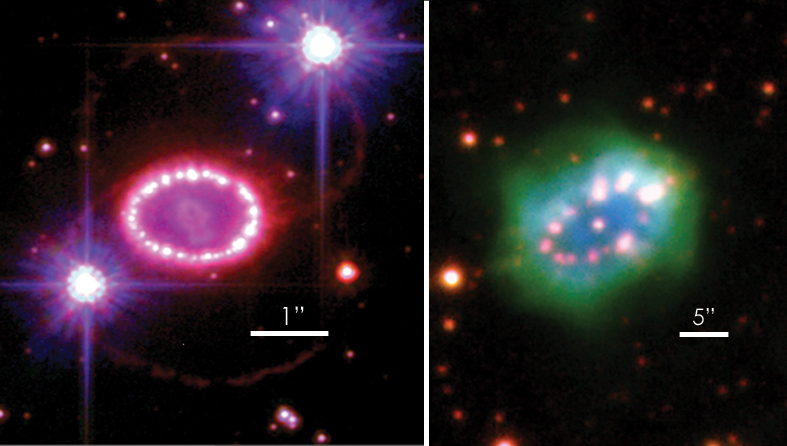}
	\caption{HST image of the ring around SN 1987A (STScI-PRC07-10; left -- Credit: NASA, ESA, P. Challis, and R. Krishner) compared with a Nordic Optical Telescope image of the Necklace planetary nebula (right -- Credit: \citet{corradi_2010}). Note the different scales indicated by white bars.\label{fig:sn87a}}
\end{figure}

Imaging on 8-m class telescopes also allowed them to derive for the first time close binary induced morphological trends. Canonical bipolar nebulae, low-ionisation structures, particularly in ring configurations, and polar outflows or jets are all provisionally associated with binarity. This correlation could fast track the discovery of more binary central stars in PNe and thereby reveal their role in the shaping of, and possibly formation of, the surrounding nebulae. 
One of the most interesting of these distinctive features is the presence of knotty rings slowly expanding and better seen in low excitation lines, such as [NII]. Such rings have already been seen in NGC 6337, Sab41, and especially the Necklace nebula (IPHASXJ194359.5+170901), which was recently shown by \citet{corradi_2010} to contain a binary CSPN with an orbital period of 1.16 days. Such a ring is very reminiscent of the one seen in SN1987A (Fig.~\ref{fig:sn87a}) and similar to those around massive stars \citep{smith_2007}.

%
%

\section{Understanding common-envelope evolution}
Post-common envelope (CE) central stars of planetary nebulae (CSPN) are an incredible, untapped tool to impose constraints on the CE interaction \citep{ibenlivio1993}. These CSPN binaries are surrounded by the ejected CE (the nebula) whose short lifetime guarantees the central binary is ``fresh out of the oven''. 
The common envelope process is unfortunately one of the least well understood stage of binary system evolution and mostly relies on a parametric formalism. It is still  a hotly debated topic \citep{ivanova_2011}, and for now is confined to parametric considerations (based on either energy or momentum transfer between the binary and envelope, neither of which appear to be particularly satisfactory). This is particularly unfortunate as increased understanding of the CE process has many applications, e.g. the role of SN Ia explosions in the chemical evolution of galaxies and their use as standard candles in cosmology \citep{branch_1993}. The CE ejection efficiency is a key parameter in
population synthesis models that has a dramatic influence on the
expected orbital period distribution, space density, birthrate and
types of post-CE binaries \citep{davis_2010}.
Given the theoretical uncertainty surrounding the efficiency parameter
\citep{politano_2007}, it is crucial to constrain
this quantity observationally by directly measuring masses and orbital
periods for a representative post-CE population. 

Close binary central stars of PN are not only a ``fresh'' post-CE population with a period unaltered by further angular momentum loss mechanisms \citep{schreiber_2003}, but they are also almost guaranteed to derive from a CE that occurred on the AGB. With a much better populated orbital period distribution, \citet{miszalski_2009a} compared the observed period distribution with predictions made by CE population synthesis models. They found that the best model relied on assuming that the secondary was drawn from a stellar Initial Mass Function, instead of the usually used constant mass ratio distribution. Another major result is the independent finding that post-CE population synthesis models overpredict the number of systems with long periods compared to
the observed period distribution, in agreement with other studies \citep{rebassa-mansergas_2008, davis_2010}.

In order to better constrain the models, it is imperative to increase by at least another factor of two the known sample of close binary CSPN. This search for new binaries can be fast-tracked by using the several morphological traits linked to binarity that were found by \citet{miszalski_2009b}. As part of a larger team, we have thus embarked in a large effort to find new binaries using a variety of telescopes. Three recently found binaries are summarised in the next section.
In parallel to this effort it is also crucial to measure the binary fraction at long orbital periods. Such systems could be the progeny of the binary post-AGB with orbital periods in the range 120--1800 days \citep{vanwinckel_2007} and the progenitors of Barium stars. In Sect.~\ref{sec:a70}, we present such a newly discovered system.

\section{New close binary CSPNe}\label{sec:new}
We discovered that the newly found ETHOS 1 (PN G068.1+11.1) is an irradiated close binary central star with an orbital period of 0.535 days and an amplitude of 0.816 mag \citep{ethos1_2011}. The extreme amplitude is consistent with the presence of a very hot central star that produces the highly ionised nebula ($T_{\rm eff} = 17,700$ K). The Necklace \citep{corradi_2010} and K 1-2 \citep{exter_2003} also share similarly large amplitudes and high-ionisation nebulae, although the absence of low-ionisation filaments in ETHOS 1 may suggest a slightly different evolutionary history. VLT FORS spectroscopy of the CSPN confirms the presence of a close binary with a large velocity separation between primary and secondary components and the nebula. The presence of \ion{N}{III}, C III and C IV emission lines continues the trend seen in other irradiated close binary CSPN. These weak emission lines are typical of many CSPN mistakenly classified as {\it wels} in the literature and we expect many of these will turn out to be close binaries. Further observations are required to constrain the orbital inclination, masses and radii of the binary CSPN.

%
\begin{figure}[t]
	\centering
	\includegraphics[width=0.7\textwidth]{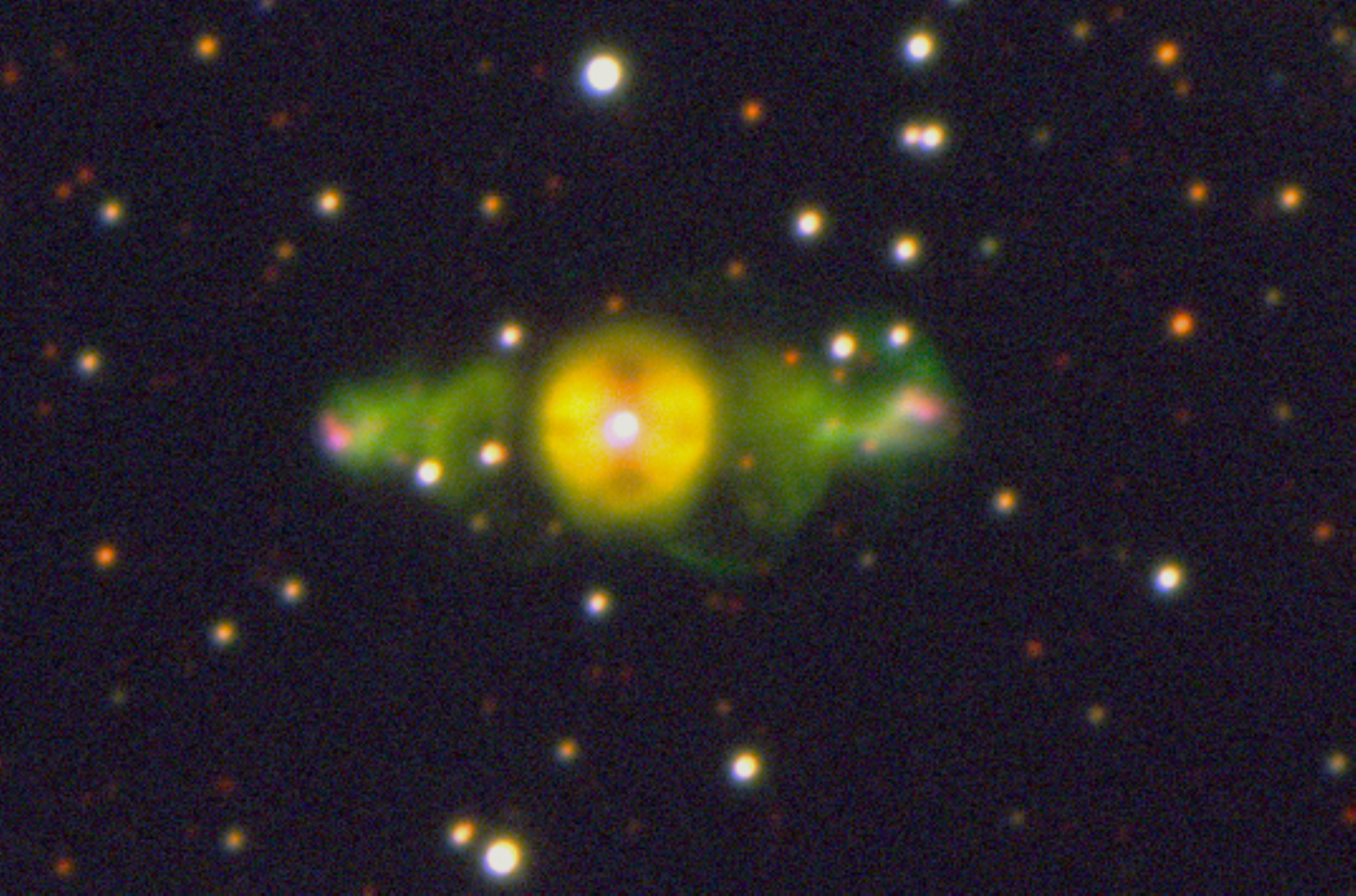}
\caption{Colour-composite image of ETHOS 1 as obtained with FORS2 on the VLT and made from images taken with H$\alpha$+[N II] (red), [O III] (green) and [O II] (blue) filters (see \citet{ethos1_2011}). \label{fig:ethos1}}
\end{figure}
%

A spectacular pair of jets travelling at 120 $\pm$ 10 km/s is found in our images. The jets are detached from the inner nebula of ETHOS 1 (Fig.~\ref{fig:ethos1}), which suggests a limited period of jet activity consistent with a transient accretion disc before the CE is ejected. The kinematic age of the jets (1750 $\pm$ 250 yrs/kpc) was found to be older than the inner nebula (900 $\pm$ 100 yrs/kpc) supporting this hypothesis. ETHOS 1 therefore continues to follow this trend previously identified in A 63 \citep{mitchell_2007} and the Necklace \citep{corradi_2010}.

We have obtained central star spectroscopy and lightcurves of the bright Southern PNe NGC 6326 and NGC 6778 that proves their close binary nature with main-sequence companions in orbital periods of 0.372 and 0.1534 days, respectively \citep{ngc6326_2011}. The combination of low-ionisation filaments and multiple jets in their nebulae further supports the suspected association between these features and post-CE binaries mentioned above. A wealth of low-ionisation filaments in NGC 6326 in particular is comparable to filaments in A 30 and A 78, which adds circumstantial evidence to the case for at least A 30 having a post-CE CSPN \citep{lau_2011}. The eclipsing CSPN of NGC 6778 is consistent with the nebula inclination determined by \citet{maestro_2004} and has one of the shortest orbital periods known for PNe.

Given absorption lines are susceptible to intrinsic non-orbital variability \citep{demarco_2009}, these three objects confirm again that the C III and N III emission lines are a powerful alternative window into detecting binarity in PNe. Their large RV shifts of 100 km/s, relative to nebula lines, make them accessible to 8-m class telescopes with intermediate resolution spectrographs. On smaller telescopes their strength means lower resolution time-domain spectroscopy could potentially reveal the tips of these lines appearing and disappearing. Such surveys are ideally suited to pre-select binaries for more time consuming photometric monitoring and will be able to routinely access CSPN of brighter nebulae that are underrepresented in known post-CE nebulae.

\section{A barium-rich cool central star}\label{sec:a70}
We have discovered that the central star of Abell 70 (A70; PN G038.1-25.4) is a G8IV/V subgiant (Miszalski et al., submitted). The radial velocity of the cool star matches the nebula emission lines, confirming that the cool star is physically associated with the nebula and we rule out a chance superposition. Moreover, the spectral energy distribution of the object reveals a clear UV-excess from GALEX that leaves no doubt on the binary nature of A70. Interestingly, we also found that the cool star presents strong lines of barium $\lambda$4554 and $\lambda$6497 due to an overabundance of s-process elements. We find [Ba/Fe]$\simeq 0.5$ from a preliminary spectral synthesis analysis of the low resolution spectra. This discovery firmly places A70 amongst the small class of barium-enhanced central stars that include only A35, LoTr 5, and WeBo 1 \citep{the_1997, bond_2003}, and as such provides a crucial link to the barium stars and to symbiotic stars, several of which have been shown to share the barium star syndrome \citep{smith_1996, smith_1997}. 

Barium stars are peculiar red giants characterised by an overabundance of carbon and s-process heavy elements. The discovery that these stars were all binaries with white dwarf companions has led to the canonical model for the formation of these stars \citep{BoJo_1988}. In this scenario the barium star was polluted -- in most cases while still on the main sequence -- by the wind of its companion that dredged up carbon and s-process elements during thermal pulses on the AGB. After ejecting its envelope as a PN, the AGB star then evolved into a white dwarf, while the contaminated star presents chemical anomalies: a barium star. 

Because of its evolutionary status -- having just left the main sequence and not yet gone through the first dredge-up that might dilute the accreted material -- the G8IV/V secondary in A70 presents us with a formidable opportunity to study several important aspects related to barium stars and planetary nebulae, in particular the s-process mechanism occurring in thermal-pulsing AGB stars, the common envelope phase as well as mass transfer and wind accretion, all of which are still very far from being understood. We therefore are planning a detailed follow-up of this ``Rosetta stone''.

\acknowledgements HB would like to thank Prof. Agn\`es Acker who introduced him to this field. The authors warmly acknowledge their collaborators over the years.

\end{document}